\documentclass[aps,twocolumn,showpacs,amsmath,amssymb,prb,superscriptaddress]{revtex4-1}
\usepackage{graphicx}
\usepackage{color}
\usepackage{amsmath}
\usepackage{psfrag}
\usepackage{natbib}

\begin{document}
\newcommand\ket[1]{|#1\rangle}
\newcommand\bra[1]{\langle#1|}
\newcommand\braket[2]{\left\langle#1\left|#2\right.\right\rangle}

\title{Tunneling-induced renormalization in interacting quantum dots} 

\author{Janine Splettstoesser}
\affiliation{Institut f\"ur Theorie der Statistischen Physik, RWTH Aachen University, D-52056 Aachen, Germany}
\affiliation{ JARA - Fundamentals of Future Information Technology}

\author{Michele Governale}
\affiliation{School of Physical and Chemical Sciences and MacDiarmid Institute for Advanced Materials and Nanotechnology, Victoria University of Wellington, PO Box 600, Wellington 6140, New Zealand}

\author{J\"urgen K\"onig}
\affiliation{Theoretische Physik, Universit\"at Duisburg-Essen and  CENIDE, D-47048 Duisburg, Germany}

\date{\today}

\begin{abstract}

We analyze tunneling-induced quantum fluctuations in a single-level quantum dot with arbitrarily strong onsite Coulomb interaction, generating cotunneling processes and renormalizing system parameters.
For a perturbative analysis of these quantum fluctuations, we remove off-shell parts of the Hamiltonian via a canonical transformation. 
We find that the tunnel couplings for the transitions connecting empty and single occupation respectively single and double occupation of the dot renormalize with the same magnitude but with \textit{opposite} signs. This has an important impact on the shape of the renormalization extracted for example from the conductance.
Finally, we verify the compatibility of our results with a systematic second-order perturbation expansion of the linear conductance performed within a diagrammatic real-time approach. 
 
\end{abstract}
\maketitle

\section{Introduction}

Few-electron quantum dots are paradigmatic systems to investigate the effects of Coulomb interaction and quantum fluctuation in nanoscale systems.~\cite{Tarucha96,Glazman03,Schon98,Andergassen10,Meir92} The theorist's workhorse in this field is the so-called Anderson-impurity model, consisting of a single spin-degenerate level with onsite Coulomb repulsion $U$, tunnel coupled to non-interacting leads. In this type of systems, the energy scale associated with Coulomb repulsion is usually large and consequently interaction effects cannot be treated within a perturbative scheme. However, when the tunnel coupling between dot and leads is smaller than temperature or voltage, a perturbation theory in the tunnel coupling strength, here denoted by $\Gamma$, can be successfully employed. 
Transport in lowest order (sometimes referred to as the sequential tunneling) can in many cases be understood by a straightforward master-equation approach with transition rates computed by means of Fermi's golden rule,~\cite{Ingold92} see for example Ref.~\onlinecite{Bruus04}. 
In this so-called orthodox theory, energy conservation allows only for transitions between energetically degenerate states, and we refer to them as classical or on-shell transitions.

Higher-order transport corrections are associated with quantum fluctuations or off-shell transitions.
In second order in $\Gamma$, there are two different types of quantum-fluctuation corrections. 
First, there are cotunneling processes.
They consist of a sequence of two tunneling events with an intermediate virtual state (i.e., the energy of the intermediate state is not equal to the energy of the initial and final state). 
Cotunneling dominates transport in the Coulomb-blockaded regions.~\cite{Beenakker91, vanHouten92} 
Some of the cotunneling processes, e.g. spin flips, occur already at zero voltage, other effects  were predicted to contribute at large bias voltage, e.g.  inelastic cotunneling~\cite{Glazman03} and pair tunneling.~\cite{Leijnse09,Koch06,Leijnse08}
The second type of quantum-fluctuation corrections are 
described by renormalizations of the system parameters, namely the dot energies and the tunnel-coupling strengths.
They lead to shifted positions and modified tunneling strengths at the transport resonances as a function of gate voltage.
In many cases, these renormalizations yield only small corrections to the lowest-order contribution to, e.g., the conductance.
There are, however, also scenarios in which the renormalization corrections are of crucial importance, because the lowest-order contribution either vanishes (as it is, e.g., the case for certain schemes in adiabatic pumping through single-level quantum dots~\cite{Splett06,Riwar10}) or provides a flat background only (as, e.g., in the maximal linear conductance through a metallic single-electron transistor, where a logarithmically temperature-dependent conductance indicated multichannel Kondo behavior~\cite{Konig97,*Konig98,Joyez97}).

Any theoretical approach that takes systematically all contributions to a given order in perturbation expansion into account will automatically include both cotunneling and renormalization effects, and, thus, go beyond pure cotunneling as obtained from standard second-order perturbation theory.~\cite{Ingold92,Sakurai94} 
This is true, e.g., for the diagrammatic real-time approach, see Refs.~\onlinecite{Konig96a,*Konig96b,*Konig99}, that was developed to describe transport through quantum-dot systems. 
In this case, however, the renormalizations can only be read off a posteriori by bringing the final result of the considered quantity (e.g., the linear conductance) into a form that allows for an identification of the renormalizations. 
We show in this paper that the linear conductance, as one possible measurable quantity to be considered, reveals these renormalization effects; however, care has to be taken, since an identification relying only on the conductance may not be unique.
Therefore, we introduce in this paper an a priori procedure to classify and evaluate the various quantum-fluctuation corrections.
This procedure is based on a canonical transformation that removes off-shell parts of the Hamiltonian and, simultaneously, generates new transitions describing cotunneling as well as renormalization of the system parameters. The main emphasis of this paper is on the derivation of the explicit expression for the renormalization of the tunnel coupling strength.

\section{Model and Method}

As a specific example, we consider a single-level quantum dot with Coulomb interaction weakly coupled to two reservoirs. Since we are interested in the linear conductance, a small bias voltage can be applied between the two reservoirs.
The system consisting of quantum dot and leads is described by the Hamiltonian
\begin{equation}
H=H_\mathrm{dot}+H_\mathrm{tunnel}+H_\mathrm{lead}\ .
\end{equation}
The single-particle level spacing in the dot is assumed to be larger than any other energy scale (temperature, Coulomb interaction, transport voltage) such that only one energy level needs to be taken into account. Hence, the dot can be  described by the single-level Anderson model 
\begin{equation}
H_\mathrm{dot}=\sum_{\sigma=\uparrow,\downarrow}\epsilon d^{\dagger}_\sigma d^{}_\sigma
+U n_\uparrow n_\downarrow\ . 
\end{equation}
The creation (annihilation) operator for an electron  with spin $\sigma$ on the dot is given by $d^{\dagger}_\sigma(d^{}_\sigma)$,  and $n_\sigma=d^\dagger_\sigma d_\sigma$ is the corresponding number operator. The onsite repulsion $U$ (as found from the constant interaction model~\cite{Glazman03}) describes the energy cost for double occupation and stems from Coulomb interaction.
Tunneling of electrons between dot and leads is taken into account  by $H_\mathrm{tunnel} =\sum_{\alpha ,k,\sigma} V_\alpha c^{\dagger}_{\alpha ,k,\sigma}d^{}_\sigma +\mathrm{h.c.}$ We assume a momentum- and spin-independent tunnel matrix element $V_\alpha$ and define the creation (annihilation) operators $c^{\dagger}_{\alpha ,k,\sigma}(c^{}_{\alpha ,k,\sigma})$ for electrons with spin $\sigma$ and 
momentum $k$ in lead $\alpha=\mathrm{L,R}$. The leads' Hamiltonian is given by $H_\mathrm{leads}=\sum_{\alpha,k,\sigma}\epsilon_{\alpha,k} c^{\dagger}_{\alpha,k,\sigma} c^{}_{\alpha,k,\sigma}$.
The chemical potential of the two leads differs by the applied bias $\mu_{L}-\mu_{R}=-eV$, with $e>0$ being the electron charge.
We assume that the density of states $\rho_\alpha$ in the leads is constant in the window relevant for transport and define the tunnel coupling strength $\Gamma_\alpha$ as $\Gamma_\alpha=2\pi\rho_\alpha|V_\alpha|^2$ and $\Gamma=\Gamma_\mathrm{L}+\Gamma_\mathrm{R}$.  

The (reduced) Hilbert space for the quantum dot is spanned by the states $|0\rangle$ for an empty dot, $|\sigma\rangle$ for a singly-occupied dot with spin $\sigma=\uparrow,\downarrow$, and $|\mathrm{d}\rangle$ for a doubly-occupied dot.
The corresponding energies are $E_0$, $E_\sigma$, and $E_\mathrm{d}$. 
The (high-dimensional) Hilbert space of the \textit{full} problem, on the other hand, is spanned by the many-body eigenstates $|n\rangle$ of the dot decoupled from the leads \textit{and} of the leads, with energy $E_n$ (containing both the energies of the lead and the dot electrons).

\section{Canonical Transformation}\label{sec_trafo}

For a systematic analysis of quantum-fluctuation effects due to tunneling, we split the Hamiltonian into three parts; a term in the absence of tunnel coupling, $H_0$, and two different types of tunneling, $H_1$ and $H_2$,
\begin{equation}
	H = H_0 + H_1 + H_2 \, .
\end{equation} 
The dot and the reservoirs in the absence of tunneling are described by 
\begin{equation}
\label{H_0}
	H_0= \sum_n E_n |n\rangle \langle n| \, .
\end{equation}
The tunneling part of the Hamiltonian introduces couplings between different eigenstates $|n\rangle$ and $|n'\rangle$.
We distinguish on-shell contributions,
\begin{equation}
\label{H_1}
	H_1 = \sum_{nn'}  V_{n'n} | n' \rangle \langle n| \delta_{E_n,E_{n'}}     \; ,
\end{equation}
which couple states of the same energy, $E_n=E_{n'}$, from the off-shell parts,
\begin{equation}
\label{H_2}
	H_2 = \sum_{nn'}  V_{n'n} | n' \rangle  \langle n| (1 - \delta_{E_n,E_{n'}}) \; ,
\end{equation}
which connect states with different energies, $E_n\neq E_{n'}$.
In both cases, we have used the abbreviation
\begin{equation}
	V_{n'n} = \langle n'| H_{\text{tunnel}} | n \rangle \, .
\end{equation}

The rates for classical (on-shell) transitions between two states $|n\rangle$ and $|n'\rangle$ are obtained via Fermi's golden rule in first order in the tunnel coupling $\Gamma$ corresponding to second order terms in $V$,
\begin{equation}\label{eq_golden}
	w_{n'n} = \frac{2\pi}{\hbar} \left| V_{n'n} \right|^2  \delta_{E_n,E_{n'}} \, ,
\end{equation}
where the Kronecker delta ensures energy conservation. 
As a consequence, only the transitions described by the on-shell part $H_1$ need to be considered. In order to find the transition rate between two states of the dot subsystem, an average over all possible initial lead states has to be performed. 
The off-shell part $H_2$ contributes to higher orders in the tunneling only.
It is this part that describes quantum fluctuations. 

The aim of this section is to remove $H_2$ and account for its effect by renormalizing the system parameters entering the on-shell part $H_0+H_1$ and by generating new transitions.
In the following, we derive this renormalization to lowest order in $\Gamma$, i.e., in second order in the tunneling matrix elements $V$.  
To this end we perform a canonical transformation~\cite{Schrieffer66,Wagner86,Winkler03,Mahan10,Bravyi11}
\begin{equation}
	\tilde H = e^{-iS} H e^{iS} 
\end{equation}
with the Hermitian operator $S$ being chosen such that $H_2$ is eliminated. 
This is achieved by $H_2 + i [H_0,S] =0$, i.e., $S$ is linear in $H_2$; the explicit form of the matrix elements of $S$ is given by $\langle m|S| n \rangle = i \langle m|H_2| n \rangle/(E_m - E_n)$, for $E_m \neq E_n$, and is equal to zero otherwise. 
With this condition, the transformed Hamiltonian $\tilde H$, expanded up to third order in the tunnel coupling $V$, reads
\begin{eqnarray}
\label{eq_htilde}
\tilde{H} & = & H_0+H_1-\frac{i}{2}\left[S,H_2\right] -i\left[S,H_1\right] -\frac{1}{2}\left[S,\left[S,H_1\right]\right] \nonumber\\
&&-\frac{1}{3}\left[S,\left[S,H_2\right]\right] \ .
\end{eqnarray}

We split $\tilde H=\tilde H_0 + \tilde H_1 + \tilde H_2$, again, into a diagonal part $\tilde H_0$, an off-diagonal but on-shell part $\tilde H_1$, and an off-shell contribution $\tilde H_2$.
The new effective model is obtained by dropping $\tilde H_2$, i.e., by dropping $-i\left[S,H_1\right]$ and all other non energy-conserving contributions which appear in Eq.~(\ref{eq_htilde}), that would contribute to processes in yet higher orders only. In the next three subsections we analyze the effect of the different corrections contributing to $\tilde{H}$.

\subsection{Energy renormalization}

The diagonal term, $\tilde H_0$, contains the renormalized energies $\tilde E_n = E_n + \delta E_n$ (for the combined system of dot plus reservoirs), with 
\begin{equation}
	\delta E_n  =  - \sum_m \frac{|\langle m|H_2|n\rangle|^2}{ E_m-E_n}  \, .\label{eq_diagren}
\end{equation}
The renormalization depends on the initial state $n$. Since we treat the leads as a reservoir in equilibrium, we average over the reservoir part of the initial states $n$ according to the Fermi distribution function 
$f_\alpha(\omega)= \{ 1+\exp [\beta(\omega-\mu_\alpha)]\}^{-1}$ where $\beta=1/(k_{\text{B}}T)$ is the inverse temperature. 
This leads to the following renormalization of the dot energies, 
\begin{eqnarray}
	\delta E_0 & = & - 2\sum_\alpha \frac{\Gamma_\alpha}{2\pi} \int d\omega \frac{f_\alpha(\omega)}{\epsilon-\omega}\label{eq_ren_E0}\\
	\delta E_\sigma & = & - \sum_\alpha \frac{\Gamma_\alpha}{2\pi} \int d\omega \left[\frac{1-f_\alpha(\omega)}{\omega-\epsilon} + \frac{f_\alpha(\omega)}{\epsilon+U-\omega} \right]\\
	\delta E_d & = & - 2\sum_\alpha \frac{\Gamma_\alpha}{2\pi} \int d\omega \frac{1-f_\alpha(\omega)}{\omega-\epsilon-U}\label{eq_ren_Ed}  \ .
\end{eqnarray}
Energy renormalizations of this type have been discussed in more complex quantum dot systems, where a non equilibrium of a spin or pseudo spin occupation can occur, see e.g. Ref.~\onlinecite{Wunsch05}. We refer to this in more detail in Section~\ref{sec_ren_firstorder}. The integrals of Eqs.~(\ref{eq_ren_E0}) to~(\ref{eq_ren_Ed}) - and the ones which will be subsequently discussed - are regularized through Cauchy's principal value (which is equivalent to adding $+i0^+$ in the denominators of the integrands and taking the real part after integration). This ad hoc regularization procedure has been addressed rigorously in the context of a T-Matrix approach in Ref.~\onlinecite{Koller10}.

The addition energies are defined as $\epsilon = E_\sigma - E_0$ and $\epsilon+U = E_d-E_\sigma$. From the expressions in Eqs.~(\ref{eq_ren_E0})-(\ref{eq_ren_Ed}) one can extract the renormalization contribution to the addition energies, namely $\delta\epsilon=   \delta E_\sigma - \delta E_0 $ and  $\delta U=\delta E_d-\delta E_\sigma-\delta\epsilon$,  which are given by
\begin{eqnarray}
	\delta \epsilon &=& - \sum_\alpha \frac{\Gamma_\alpha}{2\pi} \int d\omega \left[\frac{f_\alpha(\omega)}{\omega-\epsilon} + \frac{f_\alpha(\omega)}{\epsilon+U-\omega} \right]
	\\
	\delta U &=& 0 \, .
\end{eqnarray}
The integral can be performed analytically leading to the energy level renormalization
\begin{equation}
	\delta \epsilon = \sum_\alpha \Gamma_\alpha \left[\phi_\alpha(\epsilon+U) - \phi_\alpha(\epsilon)  \right] \, ,
\label{deps}
\end{equation}
with $\phi_\alpha(x)=\frac{1}{2\pi}\mathrm{Re}\, \Psi\left(\frac{1}{2}+i\frac{\beta (x-\mu_\alpha)}{2\pi}\right)$, where $\Psi(x)$ is the digamma function. 

\begin{figure}
\begin{center}
\includegraphics[width=3.in]{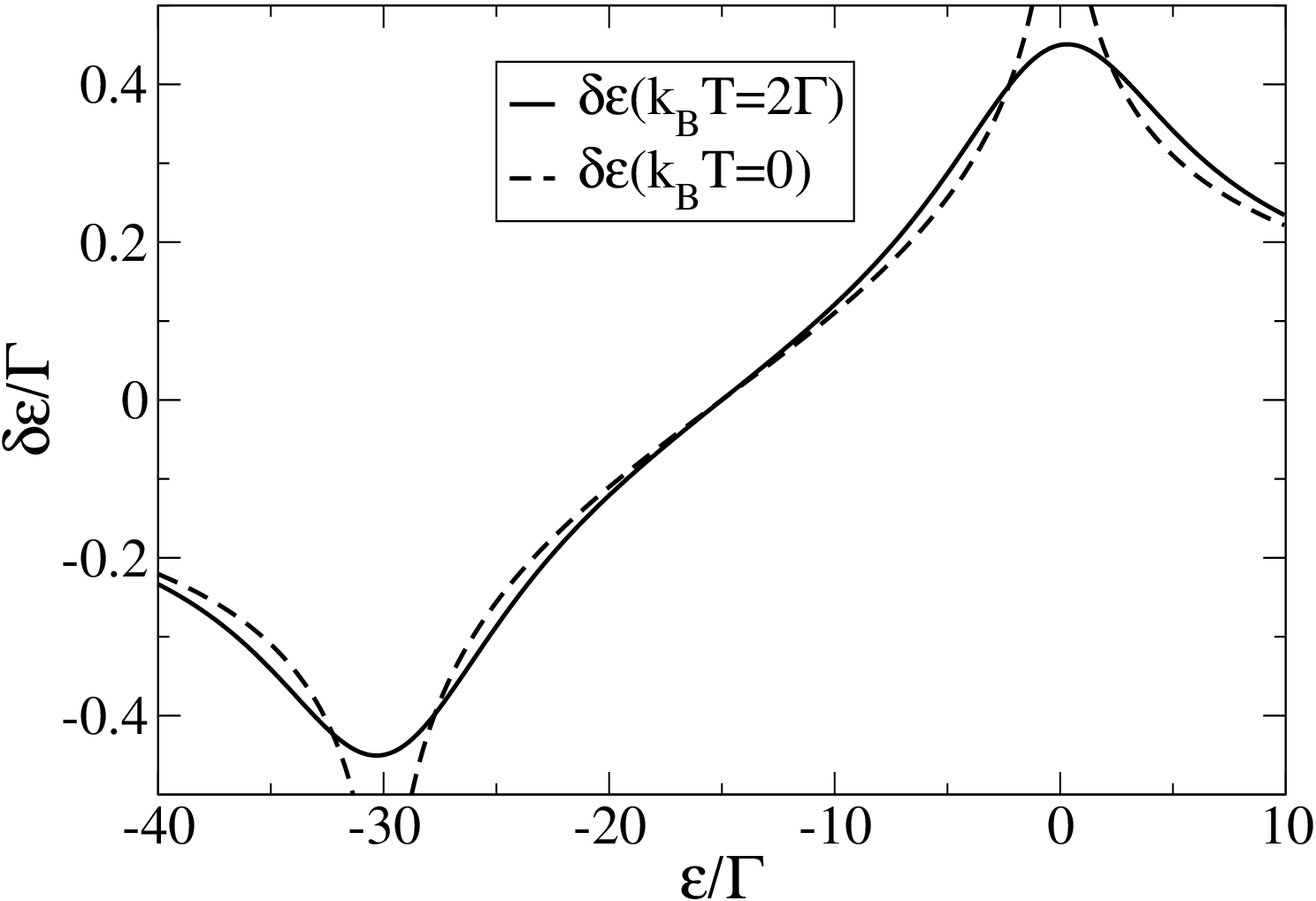}
\end{center}
\caption{Level renormalization $\delta\epsilon$  as a function of the bare level position $\epsilon$ in units of $\Gamma$, for zero bias voltage. We furthermore choose $k_\mathrm{B}T=2\Gamma$ (full line)  respectively $k_\mathrm{B}T=0$ (dashed line) and $U=30\Gamma$. \label{fig_epsren}
}
\end{figure}

The energy renormalization as function of the level position in the absence of a bias voltage is shown in Fig.~\ref{fig_epsren}.
At zero temperature, $\delta \epsilon = \frac{\Gamma}{2\pi} \ln \left| \frac{\epsilon+U}{\epsilon} \right|$ diverges logarithmically at the degeneracy points for empty and single occupation, $\epsilon=0$, and for single and doubly occupation, $\epsilon=-U$.
At finite temperature, the divergency is cut off. 
The renormalization $\delta \epsilon$ vanishes at the particle-hole symmetry point, $\epsilon=-U/2$.
Away from $\epsilon=-U/2$, the sign of $\delta \epsilon$ is such that the effective resonance positions of the system are always shifted away from the particle-hole symmetry point.

\subsection{Cotunneling}

The off-diagonal, on-shell part $\tilde H_1$ of the transformed Hamiltonian describes tunneling transitions. 
 We start with discussing the contributions occurring in second order in the tunnel matrix element $V$.
They stem from $-(i/2)\left[S,H_2\right]$ and are associated with cotunneling processes with tunnel amplitudes (with $n\neq n'$)
\begin{equation}
	\delta V_{n'n}  =  - \sum_m \frac{\langle n'|H_2|m\rangle \langle m|H_2|n\rangle}{ E_m-E_n}  \, .
\end{equation}
We remark that $n$ and $n'$ have the same parity in the dot-electron number, in contrast to the bare tunnel coupling $V_{n'n}$ that is only non-zero if $n$ and $n'$ have opposite parity.

The transition rates for these cotunneling processes are obtained by plugging the amplitudes into the Fermi golden rule, $w_{n'n} = \frac{2\pi}{\hbar} \left| \delta V_{n'n} \right|^2 \delta(E_{n'} - E_n)$, and, afterwards, averaging - as we did before - over the initial reservoir occupation.
The rates describing transitions in which the dot charge changes by $2e$ are given, in their integral form, by
\begin{eqnarray}
	w^{\alpha\alpha'}_{\mathrm{d}0} & = & 2\frac{\Gamma_\alpha\Gamma_{\alpha'}}{2\pi\hbar}\int d\omega 
	f^+_\alpha(\omega)f^+_{\alpha'}(2\epsilon+U-\omega)  \nonumber \\
	&& \times \left( \frac{1}{\omega-\epsilon} - \frac{1}{\omega-\epsilon-U} \right)^2 \ \label{eq_cot_start}
\\
	w^{\alpha\alpha'}_{0\mathrm{d}} & = & 2\frac{\Gamma_\alpha\Gamma_{\alpha'}}{2\pi\hbar}\int d\omega 	f^-_\alpha(\omega)f^-_{\alpha'}(2\epsilon+U-\omega)  \nonumber \\
	&& \times \left( \frac{1}{\omega-\epsilon} - \frac{1}{\omega-\epsilon-U} \right)^2 \ ,
\end{eqnarray}
where we used the notation $f^+_\alpha (\omega) = f_\alpha (\omega)$ and $f_\alpha^-(\omega)=1-f_\alpha(\omega)$  and $\alpha,\alpha'$ denote the leads involved in the tunneling process.
Furthermore, there are rates for processes in which the spin of the dot is flipped via an empty or doubly occupied dot as intermediate state,
\begin{equation}
	w^{\alpha\alpha'}_{\sigma\bar{\sigma}}  = \frac{\Gamma_\alpha\Gamma_{\alpha'}}{2\pi\hbar}\int d\omega f^+_{\alpha}(\omega)f^-_{{\alpha'}}(\omega) \left(\frac{1}{\omega-\epsilon}-\frac{1}{\omega-\epsilon-U}\right)^2\ . 
\end{equation}
Here, $\bar \sigma$ denotes the spin opposite to $\sigma$.
There are also rates, that leave the state of the dot unchanged.
This happens when an electron enters from one lead and leaves to possibly another one.  
The rates are given by
\begin{eqnarray}
	w^{\alpha\alpha'}_{00} & = &2 \frac{\Gamma_\alpha\Gamma_{\alpha'}}{2\pi\hbar}\int d\omega 
	\frac{ f^+_{\alpha}(\omega) f^-_{\alpha'}(\omega)}{(\omega-\epsilon)^2}\\
	w^{\alpha\alpha'}_{\sigma\sigma} & = & \frac{\Gamma_\alpha\Gamma_{\alpha'}}{2\pi\hbar}\int d\omega   f^+_{\alpha}(\omega)f^-_{\alpha'}(\omega)\nonumber\\
	&& \times\left( \frac{1}{(\omega-\epsilon)^2} 
	+ \frac{1}{(\omega-\epsilon-U)^2} \right)
 \\
	w^{\alpha\alpha'}_\mathrm{dd} & = &2  \frac{\Gamma_\alpha\Gamma_{\alpha'}}{2\pi\hbar}\int d\omega 
	\frac{ f^+_{\alpha}(\omega) f^-_{\alpha'}(\omega)}{(\omega-\epsilon-U)^2} \ .\label{eq_cot_end}
\end{eqnarray}
Again, all the integrals are regularized by adding $+i0^+$ in all resolvents and taking the real part after integration.
This integral form is convenient to trace back how many electrons have left or entered a specific lead $\alpha$: the factor $f^+_\alpha (\omega)$ indicates that an electron has left and $f^-_\alpha (\omega)$ that an electron has entered lead $\alpha$.
The analytic evaluation of the integrals is straightforward; the result is given in the appendix.

\subsection{Tunnel-coupling renormalization}

Finally, we deal with the contributions to the off-diagonal, on-shell part $\tilde H_1$, which occur in third order in $V$. These contributions renormalize the tunnel matrix elements $V_{n'n}$.
We get
\begin{eqnarray}
	 \delta V_{n'n} &=&  - \frac{\langle n'|H_1|n\rangle}{2}  \sum_{m} 
		     \frac{|\langle m|H_2|n\rangle |^2 +  |\langle n'|H_2|m\rangle|^2}{(E_{m}-E_n)^2} 
\nonumber \\
		&& + \sum_{mm'} \frac{\langle n'|H_2|m'\rangle \langle m'|H_2|m\rangle \langle m|H_2|n\rangle}
		    	{ (E_{m'}-E_n)(E_{m}-E_n)} 		
\, .
\end{eqnarray}
Again, we subsequently average over the occupation of the leads in order to get the renormalization of the tunnel matrix elements of the bare Hamiltonian, $V_\alpha$. 
Both the situations where an initially occupied or an initially empty state in one of the leads is required appear; this leads to contributions with the weight $f^+_{\alpha'} (\omega)$ and $f^-_{\alpha'} (\omega)$. 
After combining them and making use of the fact that $\int d\omega \frac{1}{(\omega-\epsilon+i0^+)^2}=0$, we find for the renormalization of the transition amplitude connecting an empty with a singly-occupied dot (with dot excitation energy $\epsilon$) 
\begin{equation}\label{eq_gammaeps}
\frac{\delta V_{\alpha,\epsilon}}{V_{\alpha}}= - \frac{1}{2} \sum_{\alpha'} \frac{\Gamma_{\alpha'}}{2\pi} \int d\omega f_{\alpha'}(\omega)
\left(\frac{1}{\omega-\epsilon} - \frac{1}{\omega-\epsilon-U}\right)^2 \, .
\end{equation}
The subscript $\epsilon$ indicates that the transition between an empty and a singly occupied dot with an excitation energy $\epsilon$ is considered here. 
The conservation of hermiticity of the canonical transformation yields 
$\delta V^*_{\alpha,\epsilon}/V^*_\alpha=\delta V_{\alpha,\epsilon}/V_\alpha$. 
Most notably this renormalization is different from the transitions connecting a singly-occupied with a doubly-occupied dot (with dot excitation energy $\epsilon+U$).
To be more specific, we find that it is equal in magnitude but has an opposite sign compared to the previous one,
\begin{equation}\label{eq_gammaepsU}
	\frac{\delta V_{\alpha,\epsilon+U}}{V_{\alpha}} = 
	\frac{\delta V^*_{\alpha,\epsilon+U}}{V^*_{\alpha}} = 
	- \frac{\delta V_{\alpha,\epsilon}}{V_{\alpha}} \, .
\end{equation}
As a consequence, the renormalization of $\Gamma_\alpha$ is given by  
\begin{eqnarray}
	\frac{\delta\Gamma_{\alpha,\epsilon}}{\Gamma_{\alpha}} =
	-\frac{\delta\Gamma_{\alpha,\epsilon+U}}{\Gamma_{\alpha}} &=&
	- \sum_{\alpha'} \Gamma_{\alpha'} \Phi_{\alpha'}(\epsilon,U)
	\ ,
	\label{dG}
\end{eqnarray}
with $\Phi_{\alpha'}(\epsilon,U)=[ \phi'_{\alpha'}(\epsilon+U)+ \phi'_{\alpha'}(\epsilon) 
	-\frac{2}{U}\{\phi_{\alpha'}(\epsilon+U) - \phi_{\alpha'}(\epsilon)\} ]$.
\begin{figure}
\begin{center}
\includegraphics[width=3.in]{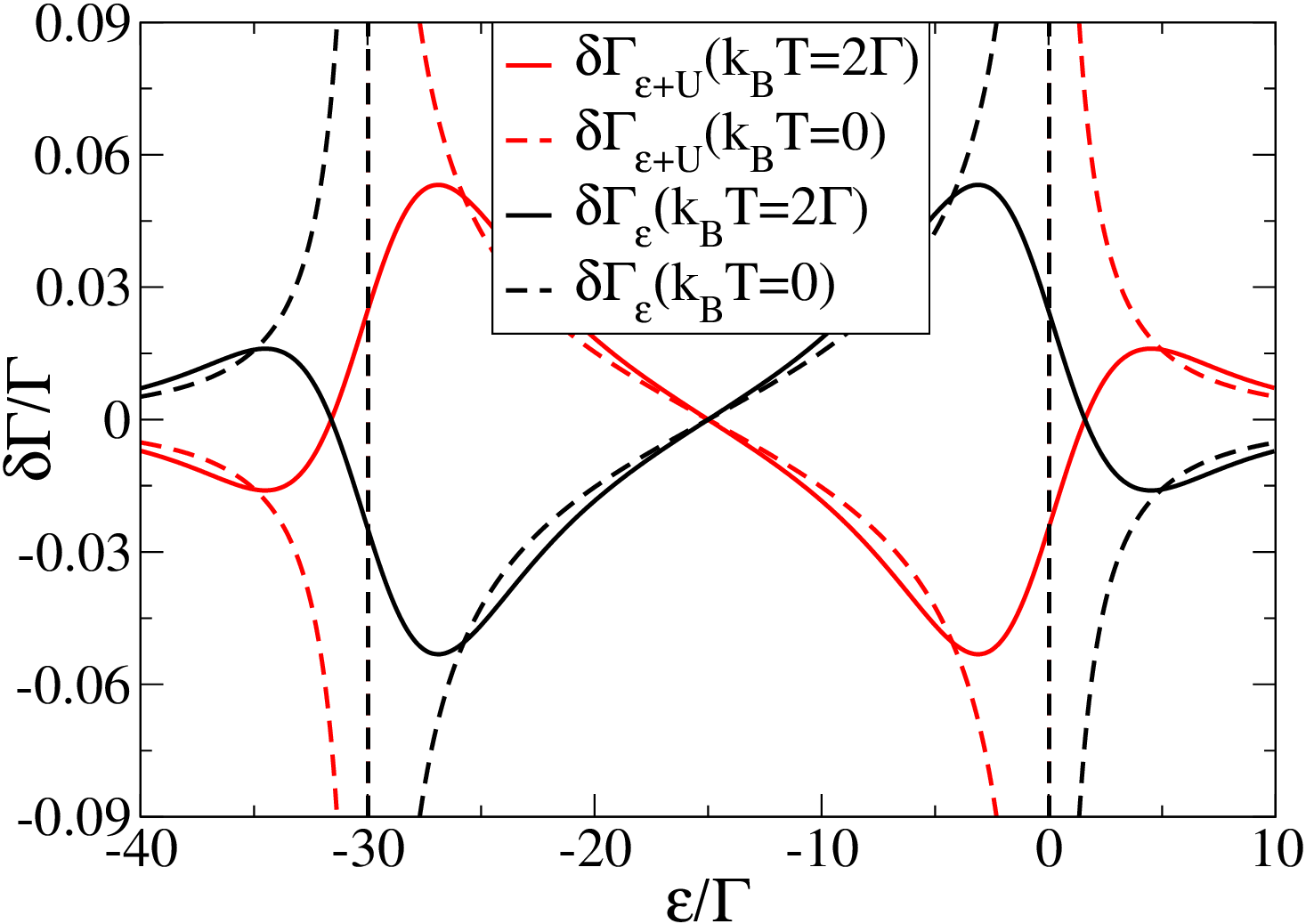}
\end{center}
\caption{(color online) Renormalization of the tunnel coupling strengths for the two transitions between empty and singly-occupied ($\delta \Gamma_\epsilon$, black) and between singly- and doubly-occupied dot ($\delta \Gamma_{\epsilon+U}$, red) as functions of the level position $\epsilon$ in units of $\Gamma$, for zero bias voltage.
We furthermore chose $k_\mathrm{B}T=2\Gamma$ (full lines) respectively $k_\mathrm{B}T=0$ (dashed lines) and $U=30\Gamma$. 
\label{fig_gammaren_sep}
}
\end{figure}
The result of Eq.~(\ref{dG})  shows that the ratio ${\delta\Gamma_{\alpha,\epsilon}}/{\Gamma_{\alpha}} $  is  the same for coupling to both leads and therefore in the figures we will suppress the index $\alpha$.  Furthermore, in the zero-bias case this ratio does not depend on the asymmetry of the bare couplings but only on $\Gamma=\Gamma_\text{L}+\Gamma_{\text{R}}$.  
The renormalization of the tunnel couplings $ \delta\Gamma_{\alpha,\epsilon}$ and $\delta\Gamma_{\alpha,\epsilon+U}$ in the absence of a bias voltage is shown in Fig.~\ref{fig_gammaren_sep}.
At zero temperature, the renormalization of tunnel couplings $\delta\Gamma_{\alpha,\epsilon} = - \frac{\Gamma_\alpha}{2\pi}\left[ \frac{1}{\epsilon+U}+\frac{1}{\epsilon}-\frac{2}{U}\ln \left| \frac{\epsilon+U}{\epsilon} \right| \right]$ displays $1/x$ divergencies at $\epsilon=0$ and $\epsilon=-U$.
At finite temperature, these divergencies are cut off.  
The sign of the renormalization is such that the excitation (either $\epsilon$ or $\epsilon+U$) that is closer to the Fermi energy of the leads acquires a stronger tunnel coupling if the dot is preferably singly occupied and a weaker one for preferred empty or double occupation.
The sign for the renormalization of the tunnel coupling associated to a certain excitation (either $\epsilon$ or $\epsilon+U$) is opposite if the \textit{other} excitation (namely $\epsilon+U$ or $\epsilon$) is close to the Fermi energy.

\subsection{Comparison with poor man's scaling}

Deriving an effective low-energy Hamiltonian with renormalized system parameters is also the central idea of renormalization group approaches.
Its simplest version, poor man's scaling, iteratively integrates out the high-energy degrees of freedom.~\cite{Hewson97,Haldane78}
It is thus possible to sum up the leading logarithmic contributions. 
However, this approach does not properly account for the subleading terms. 
Applied to the single-level Anderson impurity model for $U\gg |\epsilon|$, one obtains
\begin{equation}
	\delta \epsilon = \sum_\alpha \frac{\Gamma_\alpha}{2\pi} \ln \frac{U}{\max\{|\epsilon|, k_\mathrm{B}T \} } \, ,
\end{equation}
which qualitatively describes the correct low-temperature energy renormalization.
The exact determination of the high- and low-energy cutoff appearing in the argument of the logarithm is, of course, not possible within the poor man's scaling approach. 
Furthermore, it cannot address the tunnel-coupling renormalization at all.~\cite{Haldane78}

\section{Linear conductance}\label{sec_cond}

In the following, we show that the analysis presented above can fully explain the quantum-fluctuation corrections to the dimensionless linear conductance $g = G h/e^2$ with $G=(\partial I/\partial V)|_{V=0}$ through a single-level quantum dot.
 By using current conservation $I_{\rm L} = - I_{\rm R}$, the current can be written in a symmetric form, $I= I_{\rm L}=(\Gamma_{\rm R} I_\mathrm{L} - \Gamma_{\rm L} I_\mathrm{R})/(\Gamma_{\rm L} + \Gamma_{\rm R})$, where $I_\alpha$ is the electric current flowing from lead $\alpha$ into the dot.

\subsection{First order}

We start by considering the current $I_\alpha$ expanded to first order in the tunnel coupling. It  is obtained by evaluating the first-order tunneling rates $w_{\chi'\chi}$ that describe the change of the dot state from $\chi$ to $\chi'$, see Eq.~(\ref{eq_golden}), and multiplying them with a factor $+1$ ($-1$) when an electron enters (leaves) the dot during the transition. 
Finally, these rates multiply the (zeroth-order in the tunnel coupling $\Gamma$) probability $p_\chi(V)$ to find the dot in state $\chi \in \left\{ 0,\uparrow,\downarrow, \mathrm{d} \right\}$, when a bias voltage $V$ is applied.
Afterwards, we linearize in $V$ and arrive at the dimensionless linear conductance 
\begin{eqnarray}
\label{g1}
	g^{(1)}  &=& - \sum_{\sigma={\uparrow,\downarrow}}\left[
	\left( \frac{\Gamma_\mathrm{L}\Gamma_\mathrm{R}}{\Gamma_\mathrm{L}+\Gamma_\mathrm{R}} \right)_\epsilon f'(\epsilon) \left(p_\sigma+p_0\right) \right. \nonumber\\
& & \left.  + \left( \frac{\Gamma_\mathrm{L}\Gamma_\mathrm{R}}{\Gamma_\mathrm{L}+\Gamma_\mathrm{R}} \right)_{\epsilon+U} f'(\epsilon+U)\left(p_\sigma+p_\mathrm{d}\right) \right]\ .\label{eq_first_order}
\end{eqnarray}
Here, $p_\chi$ denotes the (zeroth-order) equilibrium ($V=0$) probability to find the dot in state $\chi$.
It is given by the Boltzmann factors $p_\chi = \exp(-\beta E_\chi) / \sum_{\chi'} \exp(-\beta E_\chi')$.
The dummy index $\epsilon$ and $\epsilon+U$ attached to the ratio of the tunnel couplings indicates the transition 
($\epsilon$ for empty/single occupation and $\epsilon+U$ for single/double occupation). 
The tunnel couplings are the same for both transitions.
However, as discussed above, they renormalize differently, which is the motivation for distinguishing them.

\subsection{Second order}

We subsequently discuss corrections to the linear conductance in second order in the tunnel coupling $\Gamma$. According to the above discussion about the different contributions to the quantum-fluctuation corrections, we express the second-order linear conductance as the sum 
\begin{equation}
	g^{(2)} = g^{(2)}_{\rm cot} + g^{(2)}_{{\rm ren},\epsilon} + \sum_\alpha \left( g^{(2)}_{{\rm ren}, \Gamma_{\alpha,\epsilon}} + g^{(2)}_{{\rm ren}, \Gamma_{\alpha,\epsilon+U}} \right) \, ,
\label{eq_second_order}
\end{equation}
containing cotunneling and different types of renormalization terms. The cotunneling part is obtained in the same way as first-order transport with the difference that cotunneling rates, see Eqs.~(\ref{eq_cot_start}) to (\ref{eq_cot_end}), instead of the sequential-tunneling rates are used.  In order to calculate the current in lead $\alpha$, these rates have again to be weighted with factors $0$, $\pm1$, respectively $\pm2$, depending on the number of particles transferred from or to lead $\alpha$ for a certain choice of $\alpha$ and $\alpha'$.
The renormalization parts are given by
\begin{equation}
	g^{(2)}_{{\rm ren},X} = \frac{\partial g^{(1)}}{\partial X} \delta X 
\end{equation}
with $X \in \{ \epsilon, \Gamma_{\alpha,\epsilon}, \Gamma_{\alpha,\epsilon+U} \}$, where we treat $\Gamma_{\alpha,\epsilon}$ and $\Gamma_{\alpha,\epsilon+U}$ as independent parameters.
The values for $\delta \epsilon$, $\delta \Gamma_{\alpha,\epsilon}$ and $\delta \Gamma_{\alpha,\epsilon+U}$ are taken from Eqs.~(\ref{deps}) and (\ref{dG}).

To verify the validity of this result, we recalculate the conductance within a diagrammatic real-time approach.~\cite{Konig96a,*Konig96b,*Konig99}
The virtue of this approach is that it allows for a systematic perturbation expansion that does not require the identification and the separation of different quantum-fluctuation contributions. 
The analytical expression that we obtain for the first- and second-order conductance within this alternative approach is identical to the one obtained from Eqs.~(\ref{eq_first_order}) and (\ref{eq_second_order}).

\begin{figure}
\includegraphics[width=3.in]{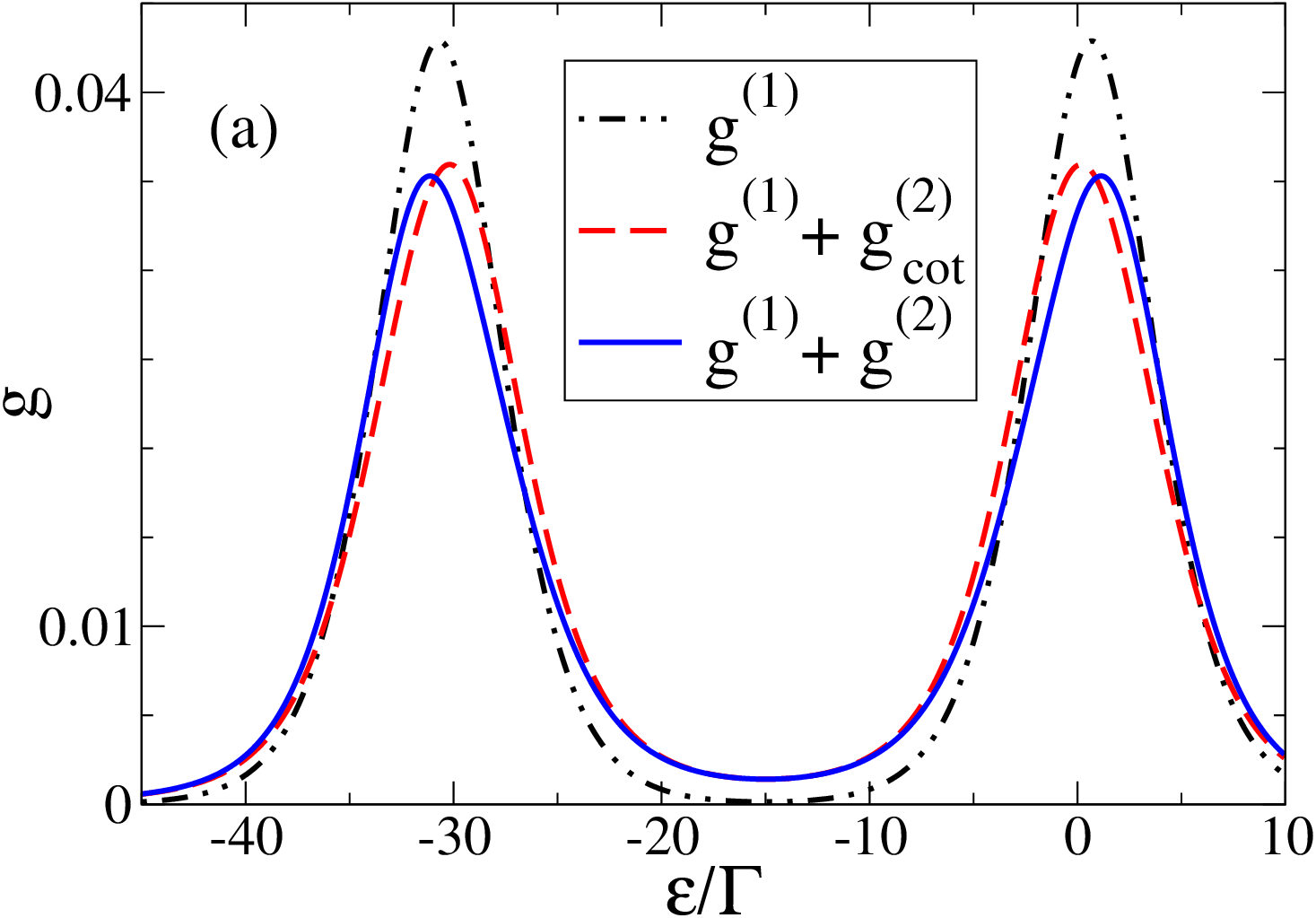}
\includegraphics[width=3.in]{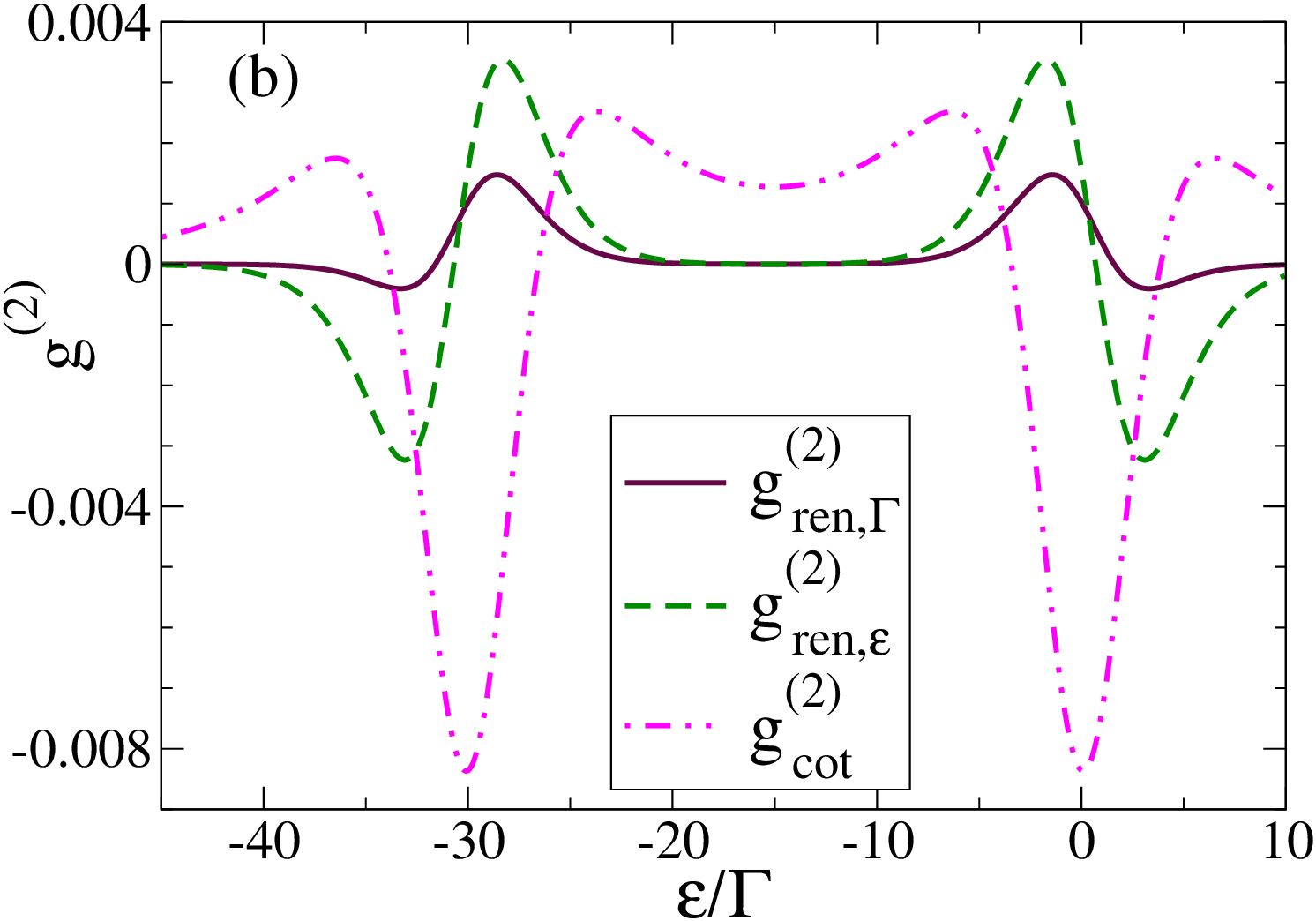}
\caption{(color online) (a) Dimensionless conductance as a function of the level position in units of $\Gamma$ in first order (black, dashed-dotted line), including second order corrections due to real cotunneling (red, dashed line), and including all second order corrections (blue, full line). (b) Separate second-order corrections: real cotunelling (pink, dashed-dotted line), level renormalization (green, dashed line) and $\Gamma$-renormalization (violet, full line). We choose $\Gamma_\mathrm{L}=\Gamma_\mathrm{R}$, $k_\mathrm{B}T=2\Gamma$ and $U=30\Gamma$.
\label{fig_cond}}
\end{figure}

The results for the conductance are plotted in Fig.~\ref{fig_cond}(a) as a function of the level position.
The black, dashed-dotted line is the conductance in first order in the tunnel coupling. As it is well known, the appearing peaks correspond to the addition energies of the dot being at resonance with the chemical potential of the leads.
The blue, full line is the full first-order plus second-order result.
For comparison, we also show (red, dashed line) the sum of sequential- and cotunneling under the neglect of renormalization corrections.
In Fig.~\ref{fig_cond}(b) only the separate second-order corrections are displayed. As expected, we observe from these figures that only cotunneling effects lead to finite contributions in the Coulomb-blockaded regions while renormalization corrections contribute to the conductance only close to the resonances.

\begin{figure}
\begin{center}
\includegraphics[width=3.in]{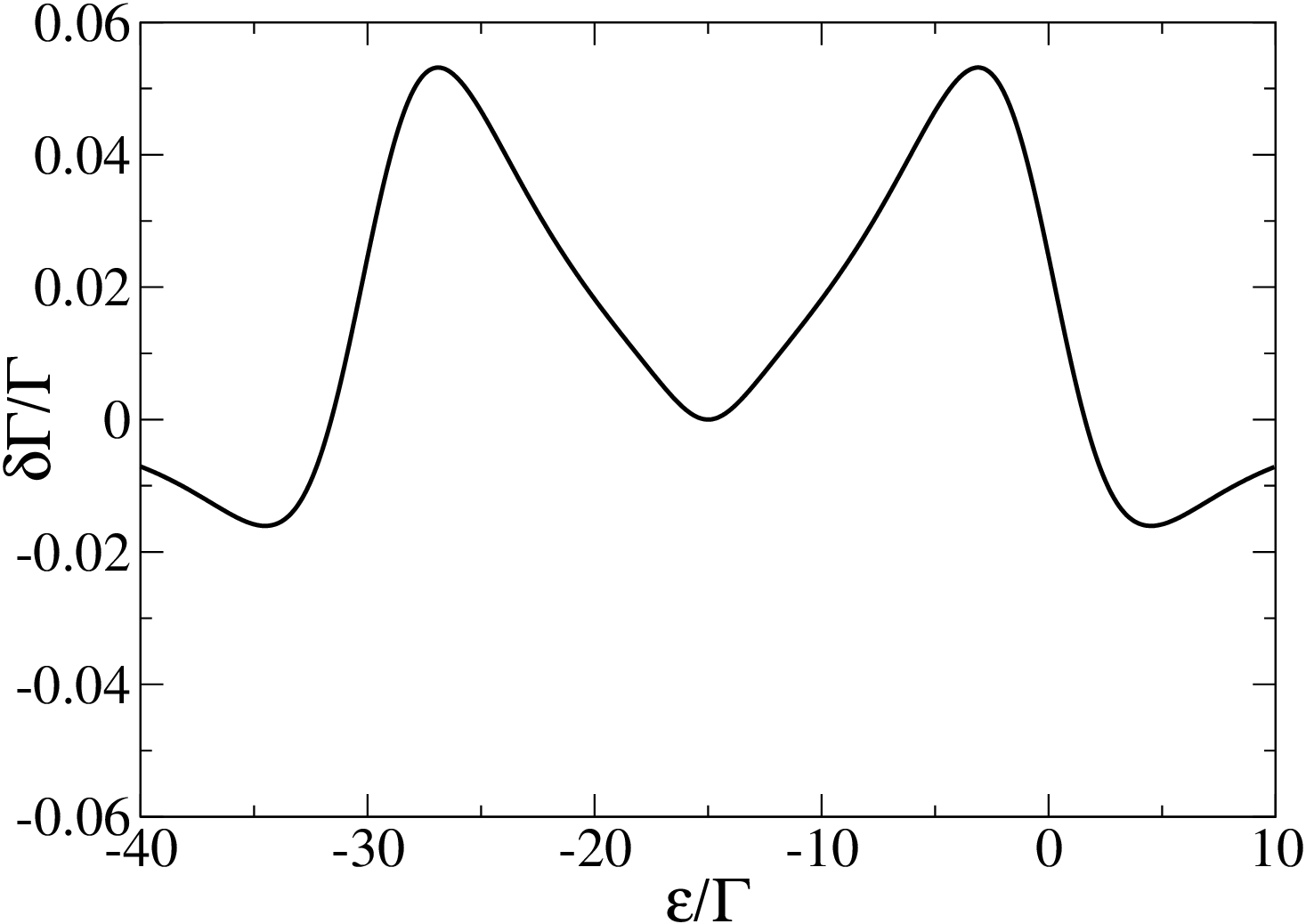}
\end{center}
\caption{Combined renormalization of the line width $\delta\Gamma$ as function of the level position $\epsilon$ in units of $\Gamma$,  for zero bias voltage.
We choose $k_\mathrm{B}T=2\Gamma$ and $U=30\Gamma$. 
\label{fig_gammaren}
}
\end{figure}

Suppose, one wants to determine the renormalizations from the analytical expression of the linear conductance calculated from the diagrammatic real-time approach directly.
Once we have written the result in the form of Eq.~(\ref{eq_second_order}), we can read off $\delta \epsilon$,  
$\delta \Gamma_{\alpha,\epsilon}$ and $\delta \Gamma_{\alpha,\epsilon+U}$.
This is, however, not the only possibility.
Supposing an overall renormalization of the tunnel coupling strength, without taking account for the  different renormalizations $\delta \Gamma_{\alpha,\epsilon}$ and $\delta \Gamma_{\alpha,\epsilon+U}$ of the two resonances, quantum-fluctuation effects on the tunnel coupling were described by one tunnel-coupling renormalization $\delta \Gamma_\alpha$ only, i.e., $g^{(2)} = g^{(2)}_{\rm cot} +  \frac{\partial g^{(1)}}{\partial \epsilon} \delta \epsilon  + \sum_\alpha \frac{\partial g^{(1)}}{\partial \Gamma_\alpha} \delta \Gamma_\alpha$.
This description  is indeed possible and leads to
\begin{equation}
\label{dGs}
	\delta \Gamma_\alpha = \frac{1-f(\epsilon)-f(\epsilon+U)}{1-f(\epsilon)+f(\epsilon+U)}  
		\delta \Gamma_{\alpha,\epsilon} \, ,
\end{equation}
with $\delta\Gamma_{\alpha,\epsilon}$ from Eq.~(\ref{dG}).
The above identity is easily understood by combining the condition 
$ \frac{\partial g^{(1)}}{\partial \Gamma_\alpha} \delta \Gamma_\alpha = \frac{\partial g^{(1)}}{\partial \Gamma_{\alpha,\epsilon}} \delta \Gamma_{\alpha,\epsilon} + \frac{\partial g^{(1)}}{\partial \Gamma_{\alpha,\epsilon+U}} \delta \Gamma_{\alpha,\epsilon+U}$
with the relation $\delta \Gamma_{\alpha,\epsilon} = -\delta \Gamma_{\alpha,\epsilon+U}$
that we found via the canonical transformation. The expression for the single tunnel-coupling renormalization Eq.~(\ref{dGs}) is plotted in Fig.~\ref{fig_gammaren} for the zero-bias case; this total result is electron-hole symmetric. Also in this case, the ratio $\delta \Gamma_\alpha/  \Gamma_\alpha$ is the same for both leads and we drop the lead index in the figure.
Far away from the particle-hole symmetry point, $\epsilon+U/2 \gg k_B T$ or $\epsilon+U/2 \ll -k_B T$, only one of the transitions between the empty and the single occupation or between the single and the double occupation plays a role. 
As a consequence, $\delta \Gamma_\alpha$ is given by the corresponding tunnel-coupling renormalization, $\delta \Gamma_{\alpha,\epsilon}$ or $\delta \Gamma_{\alpha,\epsilon+U}$, only.
Around the particle-hole symmetry point, $\delta \Gamma_\alpha$ interpolates between $\delta \Gamma_{\alpha,\epsilon}$ and $\delta \Gamma_{\alpha,\epsilon+U}$ by averaging with the proper weights. 

In Refs.~\onlinecite{Splett10a,Contreras12}, we calculated the relaxation rates of a quantum dot brought out of equilibrium, performing a perturbation expansion within a real-time diagrammatic approach.
Similar as for the linear conductance discussed above, the second-order corrections could be fully understood in terms of cotunneling processes and renormalization of energy and tunnel coupling. 
Ignoring the possibility of different renormalizations $\delta \Gamma_{\alpha,\epsilon}$ and $\delta \Gamma_{\alpha, \epsilon+U}$, we extracted from the relaxation rates an expression  for $\delta \Gamma_\alpha$ that coincides with the one obtained from the differential conductance given in Eq.~(\ref{dGs}).

\subsection{Energy renormalization in first-order transport}\label{sec_ren_firstorder}
 
For the simple single-level Anderson model considered in this paper, the energy renormalization gives rise to corrections in second-order transport but does not show up in the conductance in first order in $\Gamma$.
The reason is that in the quantity of interest, Eq.~(\ref{g1}), only the \textit{equilibrium} probabilities to zeroth order in $\Gamma$ enter: the corrections linear in $V$ happen to drop out. 
 This is, however, not always the case. 
For slightly more complex systems, tunneling-induced energy renormalizations already affect lowest-order transport.
In a single-level quantum dot attached to noncollinearly magnetized ferromagnetic leads, e.g., there is a spin-dependent energy renormalization that has been described as an effective, tunneling-induced exchange field that 
influences the dynamics of the quantum-dot spin in a way that is detectable already in the linear conductance to lowest order in the tunnel coupling.~\cite{Konig03,*Braun04}
This happens also for a metallic island coupled to noncollinearly magnetized ferromagnetic leads.~\cite{Wetzels05,*Wetzels06,*Lindebaum11}
A similar effect is found for the case of double quantum dots,~\cite{Wunsch05,*Trocha09,Riwar10} in which the two levels in the two dots define an isospin that experiences a similar pseudo exchange field. The effect of effective exchange fields due to level renormalization was also observed in carbon nanotubes with orbital-dependent tunnel couplings~\cite{Holm08,*Grap12,*Zitko12,*Koller12,*Kirsanskas12} and molecular single-electron transistors.~\cite{Darau09}
Furthermore, the first-order transport characteristics through quantum dots attached to superconducting leads~\cite{Pala07} shows features of different tunneling-induced energy renormalizations of the empty and the doubly occupied dot.

\section{Conclusion}
We present an approach to classify the effects of quantum fluctuations in quantum-dot systems within a perturbative expansion in the tunnel-coupling strength.  
This approach is based on a canonical transformation that removes off-shell parts of the Hamiltonian and, simultaneously, generates new transitions as well as renormalizes system parameters such as energy and tunnel coupling. 
We illustrate this idea for the example of a single-level Anderson impurity model.
Most notably, we find that the tunnel coupling strength for the two resonances connecting empty and single occupation respectively  single and double occupation of the quantum dot renormalize with opposite sign. The discussed effects are identified in a full second order expression for the linear conductance through the interacting quantum dot.

\acknowledgments
We benefitted from discussion with Herbert Schoeller and Michael Hell.
 Financial support from the Ministry of Innovation NRW is acknowledged.

\appendix

\section{Cotunneling: Analytic results}

In this section we give the expressions for the Fermi golden rule rates in terms of  digamma functions and their derivatives.  
 We find for the rates describing a transition  leaving the state of the quantum dot unchanged
\begin{eqnarray*}
w^{\alpha\alpha'}_{00} & = &  \frac{\Gamma_{\alpha}\Gamma_{\alpha'}}{\hbar}\frac{2}{1-e^{\beta(\mu_{\alpha'}-\mu_{\alpha})}}\left[\phi'_\alpha(\epsilon)-\phi'_{\alpha'}(\epsilon)\right]\\
w^{\alpha\alpha'}_{\sigma\sigma} & = &  \frac{\Gamma_{\alpha}\Gamma_{\alpha'}}{\hbar}\frac{1}{1-e^{\beta(\mu_{\alpha'}-\mu_{\alpha})}}\\
&&\hfill\left[\phi'_\alpha(\epsilon)+\phi'_{\alpha}(\epsilon+U)-\phi'_{\alpha'}(\epsilon)-\phi'_{\alpha'}(\epsilon+U)\right]\\
w^{\alpha\alpha'}_{\mathrm{dd}} & = &  \frac{\Gamma_{\alpha}\Gamma_{\alpha'}}{\hbar}\frac{2}{1-e^{\beta(\mu_{\alpha'}-\mu_{\alpha})}}\left[\phi'_\alpha(\epsilon+U)-\phi'_{\alpha'}(\epsilon+U)\right]\ .
\end{eqnarray*}
The spin-flip term is evaluated as
\begin{eqnarray*}
w^{\alpha\alpha'}_{\sigma\bar{\sigma}} & = & \frac{\Gamma_\alpha\Gamma_\alpha'}{\hbar}\frac{1}{1-e^{\beta(\mu_{\alpha'}-\mu_\alpha)}}\left[\Phi_\alpha(\epsilon,U)-\Phi_{\alpha'}(\epsilon,U)\right].
\end{eqnarray*}
with $\Phi_\alpha(\epsilon,U)$ defined as in Eq.~(\ref{dG}).
Finally we give explicit expressions for the cotunneling rates where the state of the dot changes by a charge $2e$,
\begin{eqnarray*}
w^{\alpha\alpha'}_{\mathrm{d}0} & = & \frac{\Gamma_\alpha\Gamma_{\alpha'}}{\hbar}\frac{2}{1-e^{\beta(2\epsilon+U-\mu_{\alpha}-\mu_{\alpha'})}}\left[\Phi_\alpha(\epsilon,U)+\Phi_{\alpha'}(\epsilon,U)\right]\nonumber\\
	w^{\alpha\alpha'}_{0\mathrm{d}} & = &   \frac{\Gamma_\alpha\Gamma_{\alpha'}}{\hbar}\frac{-2e^{\beta(2\epsilon+U-\mu_{\alpha}-\mu_{\alpha'})}}{1-e^{\beta(2\epsilon+U-\mu_{\alpha}-\mu_{\alpha'})}}\left[\Phi_{\alpha}(\epsilon,U)+\Phi_{\alpha'}(\epsilon,U)\right]\nonumber.
\end{eqnarray*}
All cotunneling terms can hence be expressed via the same digamma functions and their derivatives that appear in the renormalization contributions. In order to calculate the current or the conductance, these rates have to be weighted with the respective factor taking account for the number and direction of transferred particles and with the probability that the initial state is occupied.

\end{document}